# Experimental verification of the very strong coupling regime in a GaAs quantum well microcavity

S. Brodbeck,[1,*] S. De Liberato,[2] M. Amthor,[1] M. Klaas[1], M. Kamp,[1] L. Worschech,[1] C. Schneider,[1] and S. Höfling[1,3]

[1]*Technische Physik, Physikalisches Institut and Wilhelm Conrad Röntgen-Research Center for Complex Material Systems, Universität Würzburg, Am Hubland, D-97074 Würzburg, Germany*

[2]*School of Physics and Astronomy, University of Southampton, Southampton SO17 1BJ, United Kingdom*

[3]*SUPA, School of Physics and Astronomy, University of St. Andrews, St. Andrews, KY 16 9SS, United Kingdom*

When the coupling between light and matter becomes comparable to the energy gap between different excited states they hybridize, leading to the appearance of a rich and complex phenomenology which attracted remarkable interest in recent years. While the mixing between states with different number of excitations, so-called ultrastrong coupling regime, has been observed in various implementations, the effect of the hybridization between different single excitation states, referred to as very strong coupling regime, has remained elusive. In semiconductor quantum wells such a regime is predicted to manifest as a photon-mediated electron-hole coupling leading to different excitonic wavefunctions for the two polaritonic branches when the ratio of the coupling strength to exciton binding energy $g/E_B$ approaches unity. Here, we verify experimentally the existence of this regime in magneto-optical measurements on a microcavity with 28 GaAs quantum wells, characterized by $g/E_B \approx 0.64$, showing that the average electron-hole separation of the upper polariton is significantly increased compared to the bare quantum well exciton Bohr radius. This manifests in a diamagnetic shift around zero detuning that exceeds the shift of the lower polariton by one order of magnitude and the bare quantum well exciton diamagnetic shift by a factor of two. The lower polariton exhibits a diamagnetic shift smaller than expected from the coupling of a rigid exciton to the cavity mode which suggests more tightly bound electron-hole pairs than in the bare quantum well.

Light-matter coupling in semiconductor quantum well microcavities can be categorized into different regimes, depending on the coupling strength $g$ between cavity photons and quantum well excitons. In weak coupling, the presence of the microcavity mainly modifies the radiative decay rate of the exciton [1]. For larger coupling strengths, the strong coupling regime is reached where a reversible energy transfer between excitons and photons takes place. This manifests in the appearance of two new eigenmodes, the lower (LP) and upper polariton (UP), which are linear superpositions of the bare exciton and photon states. They are weighted by the Hopfield coefficients $X$ and $C$ which depend on $g$ and detuning $\Delta = E_C - E_X$ between exciton ($E_X$) and photon energies ($E_C$) [2,3]. At zero detuning and at zero wavevector, LP and UP are separated by the Rabi splitting $\hbar\Omega \approx 2g$. As $g$ is increased and becomes comparable to the exciton binding energy $E_B$, the light-matter coupling starts hybridizing different excitonic levels, effectively modifying the wavefunction of the electron-hole pair, in what has been named very strong coupling [4,5]. This leads to an additional, repulsive coupling term between electrons and holes for the UP, while electron-hole pairs are more tightly bound in the LP compared to the bare quantum well. Finally, if $g$ is on the order of the exciton energy $E_X$, the coupling is intense enough to hybridize states with different numbers of excitations, ushering the system into the ultrastrong coupling regime [6-9].

In this letter, we investigate an inorganic multi-quantum well microcavity in a magnetic field revealing the presence of the very strong coupling regime. While in such inorganic quantum well microcavities for the exciton-photon coupling the condition $g \ll E_X$ holds, large ratios $\gamma = g/E_B > 0.5$ are regularly achieved [10-14], implying that while the hybridization of states with different number of excitations can be safely disregarded (rotating wave approximation), the mixing of different single-exciton states should play a non-negligible role. Still there has so far been no clear experimental confirmation of the modification of electron-hole coupling in the very strong coupling regime. To calculate the polariton states for large $\gamma$, a variational treatment has been developed [4,15,16] where the polariton wavefunction is a superposition of the photon state with an effective exciton wavefunction $\phi \propto \exp(-r_{eh}/\rho)/\rho$. Here, $\rho$ is the average electron-hole separation $\rho = \langle (x_e - x_h)^2 + (y_e - y_h)^2 \rangle$, $x_{e,h}$ and $y_{e,h}$ are in-plane coordinates of the electron and hole and $\langle\ \rangle$ denotes the expectation value. The ratio $\lambda = a_B/\rho$ is then used as a variational parameter to find the polariton energies. As a result, LP and UP are characterized by different $\rho$ with $\rho_{LP} < a_B$. $\rho_{UP}$ can significantly exceed $a_B$ due to photon-mediated mixing of the exciton ground state with continuum states as the UP lies close to the quantum well bandgap for large $\gamma$. Both $\rho_{LP}$ and $\rho_{UP}$ are functions of $\Delta$ as well as $\gamma$. The strong coupling regime at small $\gamma$, where the exciton is treated as a rigid harmonic oscillator [17], is recovered by setting $\rho_{LP} = \rho_{UP} = a_B$.

Recently, the diamagnetic shift of polaritons has been proposed as a method to verify the regime of very strong coupling [16] as the diamagnetic shift of an electron-hole pair is proportional to $\rho$ [18]. Applying an external magnetic field of strength $B$ is a well-established tool for the investigation and manipulation of polaritons [19-21]. In the framework of strong coupling, the polariton energies are calculated from the Hamiltonian of two coupled oscillators [17]

$$H = \begin{pmatrix} E_X & g \\ g & E_C \end{pmatrix}. \tag{1}$$

With increasing magnetic field, the Rabi splitting increases [19,22-24] and the exciton energy exhibits a diamagnetic shift

$$\delta E_X = \kappa_X B^2, \tag{2}$$

where $\kappa_X$ is the diamagnetic coefficient of the quantum well exciton. The polariton energies as a function of magnetic field are then given by

$$E_{LP,UP}(B) = \frac{E_X + E_C}{2} + \frac{1}{2}\left(\kappa_{LP,UP}B^2 \mp \sqrt{[\hbar\Omega(B)]^2 + [\Delta - \kappa_{LP,UP}B^2]^2}\right), \tag{3}$$

where we allow for $\rho_{LP} \neq \rho_{UP}$ considering different exciton diamagnetic coefficients for LP and UP and we account for a dependence of the Rabi frequency upon the applied magnetic field. While we do not have a theory describing the detailed interplay between very strong coupling and the applied magnetic field, we expect the former to be the dominant effect, allowing us to consider a lowest order approximation in B [15,21,25]. In the absence of a magnetic field, $\rho_{LP,UP}$ in the very strong coupling regime can be calculated variationally as $\rho = a_B/\lambda$ with [4,15]

$$\lambda_\pm = 1 + \frac{\beta_\pm \gamma}{\alpha_\pm}, \tag{4}$$

where $\alpha_\pm^2 + \beta_\pm^2 = 1$ and

$$\alpha_\pm = \frac{1}{2} \pm \frac{\Delta/E_B - \gamma^2}{\sqrt{(\Delta/E_B - \gamma^2)^2 + 4\gamma^2}}. \tag{5}$$

We have measured the energies of LP and UP as a function of $B$ for two different microcavity samples. In both samples, the cavity is formed by a $\lambda/2$-wide AlAs layer surrounded by AlAs/Al$_{0.2}$Ga$_{0.8}$As distributed Bragg reflectors with 16 (20) mirror pairs in the upper (lower) reflector. The low number of mirror pairs results in moderate Q-factors around 1000 which allow for both polariton branches to be clearly resolved in reflectance for a wide range of detunings. Both samples were grown by molecular beam epitaxy on n-doped GaAs substrates. The first sample (1 QW-sample) incorporates a single 7nm wide GaAs quantum well in the center of the cavity. In the second sample (28 QW-sample), a total number of 28 GaAs quantum wells with 7nm width are placed in stacks of 4 quantum wells in the 7 central antinodes of the cavity light field [10]. AlAs barriers of 4nm width separate the quantum wells. The maximum intensity of the antinodes decreases in the mirrors, which is why the quantum wells placed outside the cavity contribute less to the total coupling strength. While all quantum wells collectively couple to the same cavity mode, this inhomogeneity does not affect the resulting polariton wavefunctions that only depends on the superradiant coupling $g$.

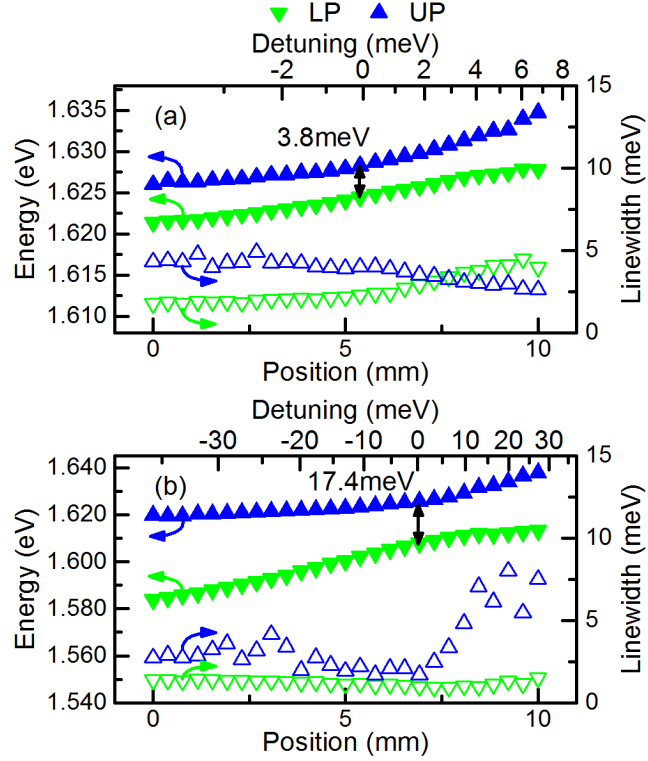

Fig. 1: Lower polariton (LP) and upper polariton (UP) energies (closed symbols) and linewidths (open symbols) as a function of sample position for samples with (a) 1 quantum well and (b) 28 quantum wells.

The Rabi splittings of both samples are determined in radial reflectance measurements at 20K where the spot of a white light source is scanned across the wafer. Because of the wedge-shape of the cavity layer introduced during the epitaxial growth, the cavity mode is tuned through the exciton resonance in these measurements. Fig. 1(a) and Fig. 1(b) show energies and linewidths of the polariton dips fitted to the reflectance spectra for the 1 QW-sample and 28 QW-sample, respectively [26]. For both samples, clear anticrossings of the cavity mode with the heavy hole-exciton are observed. The Rabi splittings amount to 3.8meV for the 1 QW-sample and 17.4meV for the 28 QW-sample, close to the highest reported value for GaAs quantum wells of 19meV for a similar sample design with 36 quantum wells [10]. With the exciton binding energy of 13.5meV for a 7nm wide GaAs quantum well in AlAs barriers [27], the ratios $g/E_B$ are approximately 0.14 (0.64) for the 1 QW-sample (28 QW-sample). For the 28 QW-sample, we observe a second anticrossing with the light hole-exciton (not shown) which lies at an energy 30meV above the heavy hole-exciton. The cavity mode could not be tuned to this energy on the 1 QW-sample. The Rabi splitting for the second anticrossing of the 28 QW-sample amounts to 15.8meV. The light hole-exciton fraction of the UP in Fig. 1(b) at zero detuning between cavity mode and heavy hole-exciton is estimated to be below 0.05. This value has been determined from modeling the coupling of the cavity mode to both excitons as coupling of three harmonic oscillators, where the coupling term between the two excitons is zero [28]. We therefore neglect the influence of the light hole-exciton in our study.

The polariton linewidths (full widths at half maximum) of the 1 QW-sample, also plotted in Fig. 1(a), show the trend expected for strong coupling. In this regime, the polariton linewidth is the average of photon and exciton linewidths which are weighted with the according Hopfield coefficients [2]. With increasing detuning the excitonic content of the LP (UP) increases (decreases). Since the exciton linewidth is larger than the photon linewidth in our samples, this results in a monotonously increasing (decreasing) linewidth for the LP (UP). Equal polariton linewidths are observed at slightly positive detuning which may be a consequence of an asymmetric exciton linewidth [29]. For the 28 QW-sample, a different behavior is observed as seen in Fig. 1(b). The LP linewidth is smaller than the UP linewidth for all detunings and reaches its smallest value (0.95meV) at zero exciton-photon detuning. The UP linewidth also decreases towards zero detuning, but then shows a drastic increase for positive detunings which is commonly observed in samples with large Rabi splittings and can be treated by including absorption by excited and continuum states in the quantum well dielectric function [10,14,30,31].

To confirm modifications of the average electron-hole separations predicted in the very strong coupling framework, we have measured diamagnetic shifts of the polariton branches in reflectance using a magneto-cryostat where the samples are held at 5K. Magnetic fields up to 5T are applied along the growth direction (Faraday configuration). The samples are illuminated by a white light source and the signal is analyzed by imaging the Fourier plane of the objective onto a spectrometer. Polariton energies are determined by fitting line spectra at $k_\parallel = 0$ with Lorentzian functions. To measure the diamagnetic shift of the uncoupled heavy hole-exciton, the upper mirrors of separate pieces of both wafers were removed by dry etching. The etched pieces are excited by a continuous wave Ti:Sapphire laser with 3mW power tuned to 1.72eV and the photoluminescence is recorded. Linear least squares fits to the diamagnetic shifts as a function of $B^2$ yield diamagnetic coefficients of $\kappa_{X,1QW} = (32.1 \pm 2.5)\mu eV/T^2$ and $\kappa_{X,28QW} = (36.7 \pm 2.8)\mu eV/T^2$ for the two samples, comparable to values measured in similar samples [21,25] The slightly smaller $\kappa_X$ of the 1 QW-sample indicates a narrower quantum well [18], in accordance with the slightly larger exciton energy of this sample, cf. Fig. 1.

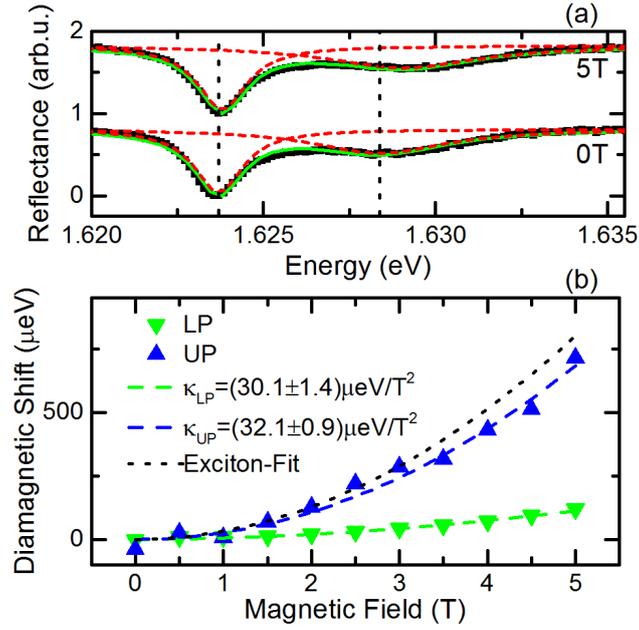

Fig. 2: (a) Reflectance spectra (black squares) at 0T and 5T for the 1 QW-sample at -3.4meV detuning. The spectra are fitted with two Lorentzian functions (red dashed lines), green solid line shows the cumulative fit. Vertical dotted lines are the fitted polariton energies at zero magnetic field. (b) Diamagnetic shifts as a function of magnetic field. Dashed lines are fits according to Eq. (3) with $\kappa$ as fitting parameter. Black dotted line is the fit to the bare exciton shift measured on a different piece of the same wafer.

Reflectance spectra at 0T and 5T for the 1 QW-sample at a detuning $\Delta = -3.4\text{meV} = -0.89\hbar\Omega$ are shown in Fig. 2(a). Both polariton dips exhibit a blueshift with increasing magnetic field. It amounts to 118μeV (714μeV) for the LP (UP) at 5T. Fitting the reflectance spectra with two Lorentzian functions yields the polariton energies as a function of magnetic field which are plotted in Fig. 2(b). The polariton energies are fitted using Eq. (3) with $\kappa_{LP,UP}$ as fitting parameters. Because of the small Rabi splitting of this sample and since an increase of only a few percent can be expected at 5T [21,25], the contribution of an increase in Rabi splitting to the energy shifts is small, see also calculations of the net diamagnetic shift for various Rabi splitting increases in the Supplementary Material [26]. We have therefore assumed a constant Rabi splitting for the fitting procedure. The fits for both polaritons yield similar values for the diamagnetic coefficients of $\kappa_{LP} = (30.1 \pm 1.4)\mu eV/T^2$ and $\kappa_{UP} = (32.1 \pm 0.9)\mu eV/T^2$ in good agreement with $\kappa_{X,1QW}$. This shows that the standard model for strong coupling which assumes rigid excitons can be applied for small $\gamma$.

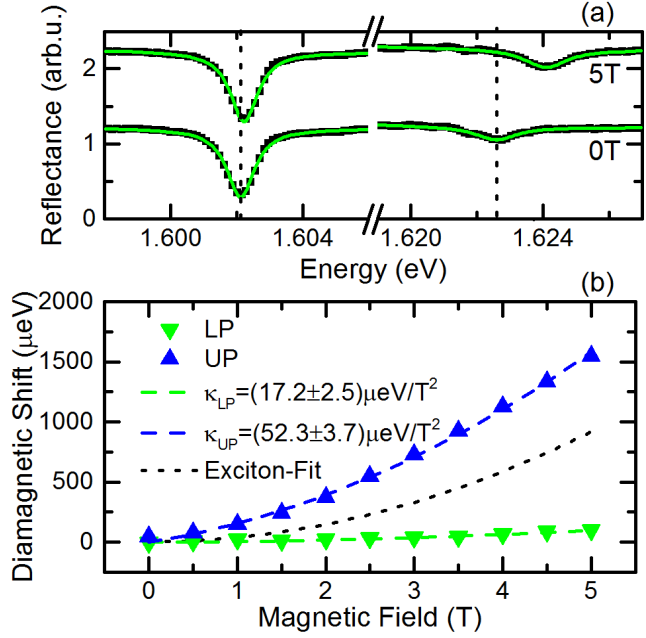

Fig. 3: (a) Reflectance spectra (black squares) at 0T and 5T for the 28 QW-sample at -10.8meV detuning. Green solid line shows the fit with two Lorentzian functions. Vertical dotted lines are the fitted polariton energies at zero magnetic field. (b) Diamagnetic shifts as a function of magnetic field. Dashed lines are fits according to Eq. (3) with $\kappa$ as fitting parameter. Black dotted line is the fit to the bare exciton shift measured on a different piece of the same wafer.

The measurement of the diamagnetic shifts for the 28 QW-sample at a detuning of $\Delta = -10.8\text{meV} = -0.62\hbar\Omega$ is presented in Figure 3. Fig. 3(a) shows reflectance spectra at 0T and 5T where again both polariton dips are shifted to higher energies with increasing magnetic field. At 5T, the diamagnetic shift of the LP amounts to 98μeV, while the shift of the UP amounts to 1.55meV and significantly exceeds the bare exciton diamagnetic shift. For this sample, the increase of the Rabi splitting with increasing magnetic field gives a larger contribution to the polariton diamagnetic shift, but this alone cannot explain the observed shifts. If we assume a typical increase of 5% at 5T as well as $\kappa_{LP} = \kappa_{UP} = \kappa_{X,28QW}$, the expected diamagnetic shifts according to Eq. (3) are 1.07meV for the UP and -153µeV for the LP which is negative in this calculation because the redshift due to an increasing Rabi splitting would exceed the blueshift due to the exciton diamagnetic shift. To fit the experimental data, we use Eq. (3) with $\hbar\Omega(B) = \hbar\Omega(0T) + cB$ and treat $\kappa$ and $c$ as free parameters with the constraints $\kappa, c \geq 0$. Fig. 3(b) depicts the diamagnetic shifts and fits as a function of magnetic field. For the LP, the best fit is achieved for $\kappa_{LP} = (17.2 \pm 2.5)$µeV/T$^2$, which is less than half as large as the bare exciton diamagnetic coefficient, and for constant Rabi splitting, i.e. $c = 0$. The fit to the UP diamagnetic shift on the other hand yields $\kappa_{UP} = (52.3 \pm 3.7)$µeV/T$^2$ and $c = (0.27 \pm 0.03)$meV/T. The large differences in fitted values for $\kappa$ and $c$ for LP and UP indicate different electron-hole separations since both the diamagnetic shift of an exciton as well as the relative increase in oscillator strength are proportional to its radius at zero field [24].

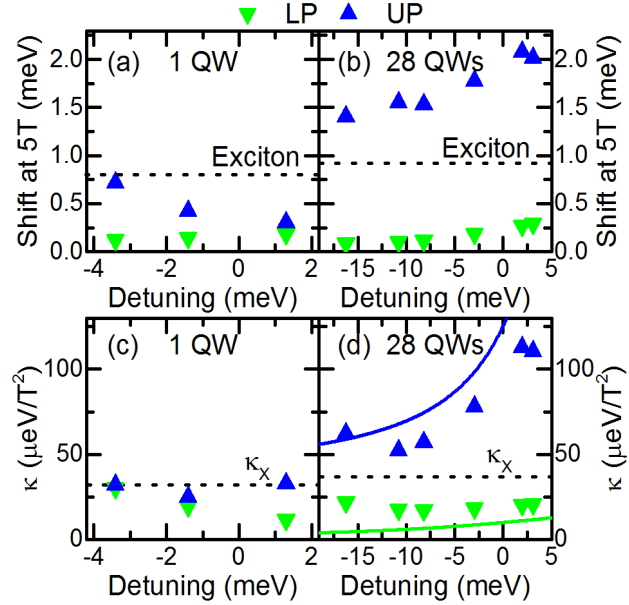

Fig. 4: Diamagnetic shifts of LP and UP at 5T for (a) the 1 QW-sample and (b) the 28 QW-sample as a function of detuning. The shift of the UP of the 28 QW-sample exceeds the bare exciton diamagnetic shift for all detunings. (c) Diamagnetic coefficients from fits according to Eq. (3) for the 1 QW-sample. (d) The same as (c) for the 28 QW-sample where $\kappa_{UP}$ exceeds $\kappa_X$ for all detunings. Solid lines are theoretical curves according to Eqs. (4) and (5) which yield $\rho_{LP,UP}$ in the very strong coupling framework.

The diamagnetic shifts of both polariton branches were measured at several different detunings for both microcavity samples. The total shifts at 5T are summarized in Fig. 4(a) for the 1 QW-sample and in Fig.

4(b) for the 28 QW-sample. The LP exhibits similar shifts for both samples which increase with increasing detuning. For the UP on the other hand, there is a significant qualitative and quantitative difference between the two samples. For the 1 QW-sample, the diamagnetic shift decreases with increasing detuning according to the decreasing excitonic content of the UP in the standard rigid exciton model. In stark contrast, the diamagnetic shift of the UP observed on the 28 QW-sample increases with increasing detuning and exceeds the bare exciton shift in each measurement. With increasing detuning, the UP energy approaches the quantum well bandgap energy which increases the contribution of continuum states. At a positive detuning of +1.9meV, we measure a diamagnetic shift of 2.08meV at 5T for the UP, twice as large as the shift of the bare exciton and nearly ten times as large as the shift of the LP for the same detuning (268µeV). The fitted diamagnetic coefficients would yield an exciton radius of 7.2nm (16.9nm) for the LP (UP) [32] which visualizes the dominant contribution of higher resonances to the UP. The diamagnetic coefficient of the bare quantum well exciton $\kappa_{X,28QW}$ corresponds to $a_B = 9.7$nm.

The diamagnetic shifts were all fitted with Eq. (3) with $\kappa$ as fitting parameter. Fig. 4(c) depicts the resulting diamagnetic coefficients for the 1 QW-sample and Fig. 4(d) for the 28 QW-sample. For the 1 QW-sample, the Rabi splitting was assumed constant for all fits. $\kappa_{LP}$ decreases with increasing detuning from 30.1µeV/T² at -3.4meV down to 11.4µeV/T² at +1.3meV. Because of the small Rabi splitting and moderate Q-factor of this sample, the splitting between LP and UP in the considered detuning range barely exceeds the linewidths of the polaritons which range from 2meV to 4meV. The LP linewidth increases with increasing detuning, Fig. 1(a), which could explain the unexpected decrease of $\kappa_{LP}$ due to greater uncertainties in the fits of the reflectance spectra. The UP diamagnetic coefficient shows no clear trend with all values in the range of $(30 \pm 4)$µeV/T² close to the bare quantum well exciton diamagnetic coefficient of this sample. For the 28 QW-sample, the polariton dips are well separated at all detunings due to the large Rabi splitting which facilitates fitting of the reflectance spectra. $\kappa_{LP}$ is in the range of $(19 \pm 2)$µeV/T² for all detunings with no clear trend visible for increasing detuning. This value is roughly half as large as $\kappa_{X,28QW}$. Additionally, the fitted values for $c$ are below 8µeV/T for all detunings which also indicates a small $\rho$. $\kappa_{UP}$ increases with increasing detuning and reaches values above 100μeV/T² for slightly positive detunings. The fitted values for $c$ also show a slight increase with detuning with values in the range of $(0.24 \pm 0.10)$meV/T. Both fit parameters are consistent with an increased $\rho$ for the UP as predicted by the framework of very strong coupling. The diamagnetic coefficients for the 28 QW-sample have also been calculated with the theoretical values for $\rho_{LP,UP}$ according to Eqs. (4) and (5) which have no free parameters. The theoretical curves for very strong coupling are shown as solid lines in Fig. 4(d) and they are in good agreement with the values determined by fitting of the diamagnetic shifts with Eq. (3). Finally, we have calculated the net diamagnetic shift at 5T for both samples in the coupled oscillator model using the measured values of $\kappa_X$ which is shown in the Supplementary Material [26]. For the 1 QW-sample, there is good qualitative and quantitative agreement of theory and experiment which shows that treating LP and UP as linear combinations of photons with a rigid exciton is a good approximation for a small Rabi splitting. For the 28 QW-sample on the other hand, the coupled oscillator model fails to reproduce the experimental values. It is essential to account for photon-mediated electron-hole coupling in this sample, e.g. by using polariton wavefunctions characterized by different $\rho$ for LP and UP.

To conclude, we have shown that coupling to a cavity mode can modify not only the radiative decay of electron-hole pairs, but may also influence their formation mechanism. The very large diamagnetic shift of the UP that we measure for a sample with $g/E_B > 0.5$ is clear evidence of an increased average electron-hole separation due to photon-mediated mixing of the optically allowed interband transitions. For the LP, the comparably small diamagnetic shift indicates a reduced electron-hole separation which is explained in the framework of very strong coupling by increased electron-hole attraction due to photon-mediated interactions. This increased attraction could be exploited to realize polariton condensates at room temperature even in semiconductors like GaAs for which the exciton binding energy of a bare quantum well is smaller than the thermal energy at room temperature [15].

We thank M. Wagenbrenner and A. Wolf for assistance during sample preparation. Helpful discussions with A. Schade and H. Suchomel are gratefully acknowledged. This work was supported by the State of Bavaria. S.D.L. acknowledges support from a Royal Society Research Fellowship and from EPSRC

Grant No. EP/M003183/1.

# References

* sebastian.brodbeck@physik.uni-wuerzburg.de